# Interplay of Kerr and Raman beam cleaning with a multimode microstructure fiber

R. DUPIOL[1,2*], K. KRUPA[3], A. TONELLO[1], M. FABERT[1], D. MODOTTO[3], S. WABNITZ[2,3,4], G. MILLOT[2], V. COUDERC[1]

[1]Université de Limoges, XLIM, UMR CNRS 7252, 123 Avenue A. Thomas, 87060 Limoges, France
[2]Université Bourgogne-Franche-Comté, ICB UMR CNRS 6303, 9 Avenue A. Savary, 21078 Dijon, France
[3]Dipartimento di Ingegneria dell'Informazione, Università di Brescia, via Branze 38, 25123 Brescia, Italy
[4]Istituto Nazionale di Ottica del Consiglio Nazionale delle Ricerche (INO-CNR), via Branze 45, 25123 Brescia, Italy
*Corresponding author: richard.dupiol@unilim.fr

**We experimentally study the competition between Kerr beam self-cleaning and Raman beam cleanup in a multimode air-silica microstructure optical fiber. Kerr beam self-cleaning of the pump is observed for a certain range of input powers only. Raman Stokes beam generation and cleanup lead to both depletion and degradation of beam quality for the pump. The interplay of modal four-wave mixing and Raman scattering in the infrared domain lead to the generation of a multimode supercontinuum ranging from 500 nm up to 1800 nm.**

http://dx.doi.org/10.1364/OL.99.099999

As well known, stimulated Raman scattering (SRS) in graded index (GRIN) fibers permits to convert a multimode, speckled pump into a high brightness, nearly single mode Stokes beam [1]. This Raman beam cleanup results from the mode selective SRS gain of GRIN fibers. On the contrary, in standard multimode step-index fibers no Raman beam cleanup of the Stokes beam into the fundamental fiber mode has been observed, since the mode selection mechanism is less efficient [2]. Recent experiments have revealed that nonlinear multimode fibers (MMF) may also be used for the delivery of high brightness laser beams [3-5]. These observations have dispelled the common belief that multimode light propagation should necessarily lead to heavily speckled, low brightness beams.

The self-induced recovery of high beam quality in GRIN MMFs is based on the mechanism of Kerr spatial beam cleaning [5]: a pump beam increases its own brightness upon propagation. Spatial beam self-cleaning has been also demonstrated in Ytterbium-doped multimode fiber with a non-parabolic refractive index profile, both in a single pass configuration [6] and within a laser cavity [7]. In GRIN MMFs, spatial beam shaping may be accompanied by a complex variety of temporal spectral shaping effects, such as multimode solitons [8], dispersive wave generation [9], intermodal modulation instabilities [10], cascaded four-wave mixing [11], geometric parametric instability [3], supercontinuum generation [12-13], and second harmonic generation [14].

To our knowledge, neither Raman beam cleanup, nor Kerr beam self-cleaning have ever been observed in a microstructure MMF. In the same vein, no supercontinuum generation has been observed in multimode microstructure fibers, although we know that such monomode fibers can generate very large supercontinua thanks to the control that we can have on its dispersive properties [15,16]. In this letter, we present the experimental observation of beam cleaning, as well as of supercontinuum generation, in a specially conceived air-silica microstructure MMF, whose transverse structure strongly differs from that of a standard weakly guiding GRIN MMF. We demonstrate that unlike for beam cleaning in GRIN fibers, the quality of the pump beam at the output of our microstructure fiber does not monotonically increase with growing input power. The Kerr-induced spatial beam compression competes with Raman cleanup. Above a certain threshold power, SRS induced Stokes beam cleaning hampers pump self-cleaning, while enhancing beam brightness on the infrared side of the supercontinuum spectrum.

Our multimode microstructure fiber is based on a hexagonal pure silica core surrounded by three layers of air holes as optical cladding, whose diameters gradually increase with the layer order (see Fig. 1(a)). The increase of the air-filling fraction when moving away from the fiber axis leads to a gradual decrease of the (azimuthal) average refractive index. Thus, the design of our microstructure fiber may be considered to be roughly equivalent to that of a standard fiber, with an intermediate index profile between that of a step-index and a GRIN fiber. However, the modes of our fiber remain remarkably different from those of a standard, weakly-guiding GRIN MMF with a parabolic profile. The diameter of the inner silica core is 30.05 µm, while the total diameter of the fiber is 145.66 µm. The guided modes and their refractive indices were computed by using COMSOL Multiphysics. Figure 1 (panels (b), (c), (d)) illustrates the numerically calculated transverse profiles and dispersion curves of the first three guided modes. In simulations made at 1064 nm, we found 55 solutions of the propagation equation in this dielectric medium that are guided modes located in the core of the fiber. At the pump wavelength, the diameter of the fundamental mode, measured at full width at half maximum in intensity (FWHMI), is 20 µm (see Fig.1 (b)). It therefore occupies all-alone two thirds of the total surface of the core of the fiber.

In our experimental setup, we used a spatially single-mode picosecond laser source at 1064 nm, delivering pulses of 60 ps with the repetition rate of 20 kHz. At the output of the laser, we placed a polarization beam splitter in between two half-wave plates, to control the power and the orientation of the linear polarization state of the incident beam. A positive lens (f = 75 mm) imaged the laser beam on the input face of the fiber with a 30 µm beam diameter and a coupling efficiency close to 30%. A micro-lens (f = 8 mm) was placed at the output end of the fiber, and we obtained the beam image, from a CCD (CMOS) camera for the spectral window of 350 nm-1150 nm (900 nm-1700 nm). The spectrum was measured by an optical spectral analyzer (OSA), Ando AQ6315E, covering the domain between 350 nm and 1750 nm, with a 0.05 nm resolution.

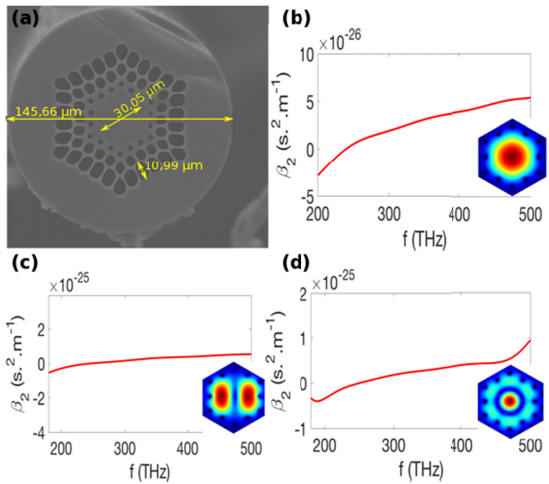

Fig. 1. (a) Cross section image of the microstructure MMF from a scanning electron microscope; numerically calculated intensity of the first three guided modes and their dispersion curves (mode 1 (b), mode 2 (c) and mode 3 (d)) computed at 1064 nm.

In a first series of experiments, we studied the power dependence of the output beam shape at both the pump wavelength (1064 nm) and at the first Raman Stokes sideband (1117 nm). These two spectral components were selected by two band-pass filters with 10 nm bandwidth each, placed in between the fiber output and the camera. The input peak power was varied by about three orders of magnitudes, from 1 kW up to about 190 kW. Figure 2 summarizes the power evolution of the transverse beam diameter, measured at the output of 11-m long fiber for both the residual pump and the first Raman Stokes beam. At relatively low input pump powers, so that beam propagation is essentially linear, the output profile at 1064 nm exhibits a typical speckled multimodal structure, with a beam diameter (at full width at half maximum in intensity, FWHMI) of 25 µm. Note that this value is close to the size of the inner silica core as it just extends over the first ring of holes. As shown by Fig.2, when increasing the input pump power above 90 kW, Kerr self-cleaning leads to a narrowing of the output beam diameter to 18 µm, (thus entirely confined in the silica core of the fiber). Moreover, when the input pump power was further increased from 20 kW up to 35 kW, the output beam profile remains substantially unchanged.

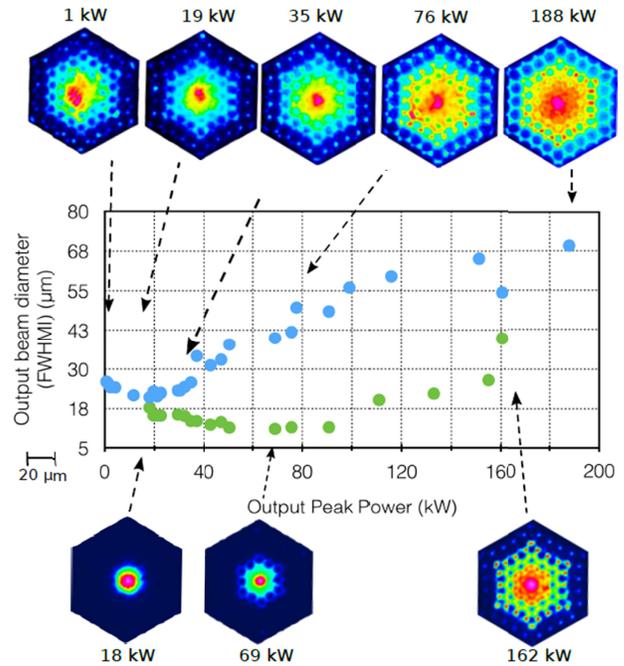

Fig. 2. Power evolution of beam diameter for the pump (blue dots) and the first Raman Stokes (green dots); insets: corresponding output beam patterns for different input peak powers for a 11-m long fiber.

Figure 2 also shows that, for input pump powers above 35 kW, the output pump beam diameter starts to grow with power. For example, for 188 kW the pump diameter has become three times wider than its low power value. Moreover, the beam shapes shown in the top insets of Fig. 2 at 76 kW and 188 kW reveal that the increase of beam diameter is associated with strong transverse beam reshaping. This can be ascribed to SRS, which introduces a strong energy depletion in the central part of the beam, while leaving nearly unchanged the outer part of the same beam (with the major contributions of high order modes). Figure 2 shows that the Stokes beam diameter (green dots) shrinks down to about 10 µm in diameter, as the input pump power grows up to 76 kW. Such a small beam diameter shows that the SRS induced frequency conversion process is accompanied by a Stokes beam cleanup, (see also the bottom insets of Fig.2, obtained for an input pump power of 18 kW and 69 kW, respectively). The relatively small diameter of the Stokes beam with respect to the fundamental mode (22 µm at the Stokes wavelength), may be explained as follows. In Ref. 2,

Terry *et al.* pointed out that Raman beam cleanup results from mode dependent gain, owing to the different overlap integrals of pump and Stokes modes. For our microstructure fiber, with a mode solver we calculated an overlap integral (that appears in the Raman gain coefficient) of 1.22 (relative to the case of both waves carried by mode 1) when the pump is in mode 1 and the Stokes is in mode 3. The overlap integral is even higher (2.82) when both the pump and the first Stokes wave are carried by mode 3. Hence the small size of the Stokes beam can be ascribed to the dominant contribution of mode 3, which has a much smaller FWHMI diameter than mode 1. When increasing the input pump power beyond 69 kW, the diameter of the Stokes also grows progressively larger with power. Note that this spatial beam broadening is accompanied by the generation of the second order Stokes wave, which depletes the first order Stokes beam. Our experiments show that SRS frequency conversion involves mode selective Raman gain, which facilitates energy conversion into the lowest order modes. As a result, SRS leads to energy drain from the high power central region of pump beams, which leads to an increase of their diameter.

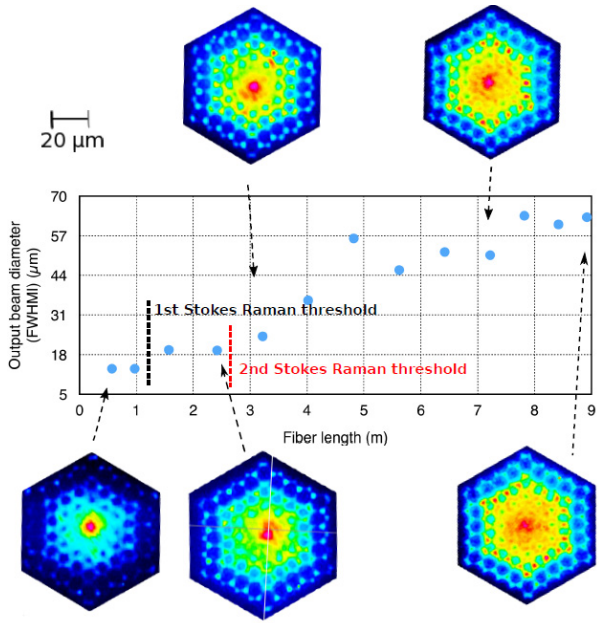

Fig. 3. Pump beam diameter vs. fiber length (input peak power: 120 kW); Insets: output pump beam pattern for different fiber lengths.

In a second series of experiments, as illustrated in Fig.3, we fixed the input peak power to 120 kW, and we measured by the cut-back method the evolution of the pump beam diameter as a function of propagation distance. For fiber lengths shorter than 1.3 m, where Raman conversion is negligible, the pump beam exhibits a clean central spot sitting on a low multimode background, a clear signature of the Kerr beam self-cleaning process. For longer propagation distances, the pump beam acquires a multimode, heavily speckled structure, with light extending up to the very last ring of holes with a diameter of about 60 μm. Strong SRS induced pump depletion quenches Kerr self-cleaning. The situation observed here is markedly different from the case of standard GRIN fibers, where Kerr self-cleaning of the residual pump beam is well preserved even after SRS-induced pump depletion [4,5]. Note also that the threshold for Kerr beam self-cleaning is relatively high with the present fiber (it is about 19 kW for a 11-m long fiber, to be compared with the 1-2 kW threshold for a 12-m long GRIN fiber [5]).

In Fig.4 we illustrate how the spectrum depends on the propagation distance for a fixed input peak power of 120 kW. As can be seen, a first nonlinear frequency conversion takes place after 0.57 m, owing to a modal four-wave mixing process (FWM) [17,18], with Stokes and anti-Stokes waves at 1277 nm and 912 nm respectively. These wavelengths agree well with the predictions of the numerically computed dispersion curves involved in the FWM phase matching condition for the modes 1 and 3 (see Fig. 1 (b,c)). These predict a 40.42 THz frequency detuning of sidebands from the pump leading to the wavelengths of 924.7 nm and 1252.7nm for the anti-Stokes and Stokes sidebands, respectively.

For fiber lengths beyond 1 m, Fig.4 shows strong SRS-induced frequency conversion into the first Stokes wave at 1117 nm. As the fiber length grows larger, we observed a progressive extension of the near-infrared spectrum (see Fig.4). For a fiber length close to 9m, a continuum ranging from 500 nm up to 1800 nm is generated. In the visible side of the spectrum, the most powerful frequency components are located at 717 nm (i.e., a 136.4 THz frequency shift from the pump), 672 nm (164.4 THz shift), and 574 nm (240.5 THz shift). All these sidebands are carried (here not shown) by higher order modes.

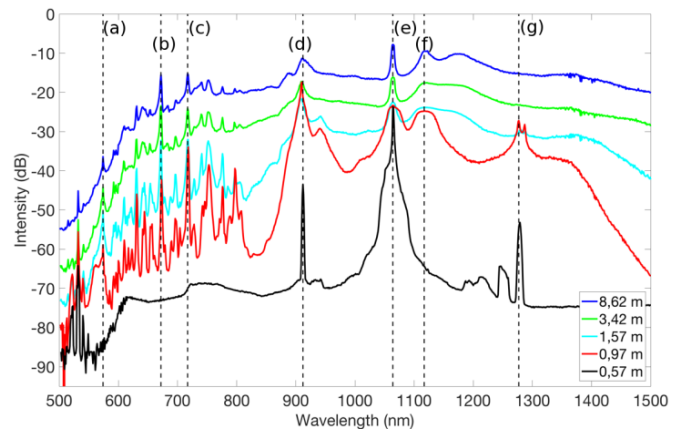

Fig. 4. Output spectrum for different fiber lengths (input peak power: 120 kW). Dashed lines correspond to the following wavelengths (detuning from the pump : (a) 574 nm (240.5 THz), (b) 672 nm (164.4 THz), (c) 717 nm (136.4 THz), (d) 912 nm (46.96 THz), (e) 1064 nm, (f) 1117 nm (13.4 THz), (g) 1277 nm (47.0 THz).

In Fig. 5 we present a comprehensive overview of the output beam shape from our 9-m long microstructure MMF, at different wavelengths and for increasing peak powers. These results were obtained by using a series of 10 nm bandpass filters in combination with the CCD and CMOS cameras, in order to cover the spectral range from 1064 nm to 1500 nm. The first row of panels (showing the power evolution of the pump beam shape) shows that Kerr beam cleaning (initially obtained for 10 kW of input peak power) is quenched by SRS, leading to a multimode speckled pattern at high powers. The second row of panels in Fig.5 shows the power dependence of the beam carried by the first-order Raman Stokes wave. Note that its generation corresponds to the beginning of the pump beam quality degradation. The Stokes beam is concentrated mainly in the all-silica core of the fiber. For input pump powers above 96 kW, the spatial quality of the first-order Stokes Raman beam starts to decrease, similarly to what was observed for the pump. In addition, we may observe that the beam shape of spectral components at 1300 nm, 1400 nm and 1500 nm closely resembles a combination of modes 1 and 3 (see Figs.1 (b,d)).

In GRIN fibers, Raman beam cleanup can be explained by the dominant modal gain of the fundamental Stokes mode, when compared with the gain of high order modes. Similarly, Raman beam cleanup observed in our microstructure MMF can be explained by the mode selection caused by the difference in overlap integrals, that appear in the exponential gain coefficient of the multimodal Raman scattering process. The continuum on the spectrum is caused by the broadening of the Raman Stokes line which is predominantly carried by the mode 3. And so, the spatial shape at the wavelengths in the infrared is close to a combination of the mode 1 and the mode 3 (see Fig. 5 at wavelengths 1300 nm, 1400 nm and 1500 nm). Note that between 1200 nm and 1300 nm, this same spatial shape should be influenced by that of the FWM peak at 1277 nm.

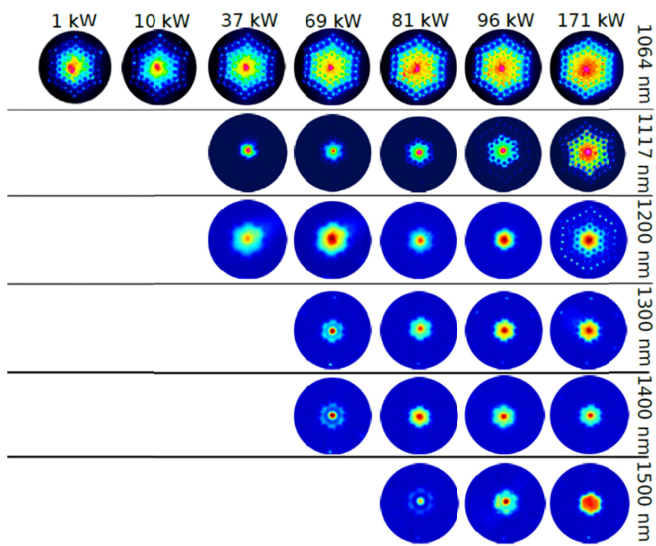

Fig. 5. Spatial evolution of the beam profile for different powers and wavelengths (fiber length of 9 m).

In conclusion, we studied nonlinear beam shaping, frequency conversion and supercontinuum generation in a specialty multimode microstructure fiber. We observed Kerr beam self-cleaning, and its competition with Raman beam cleanup. Raman conversion is particularly efficient in the fiber, which results in a progressive transfer of beam brightness towards the infrared side of the spectrum. This effect is accompanied by a loss of beam quality of the residual pump and the first Stokes wave, once it becomes a pump for higher order Stokes beams.

**Funding.** iXcore research foundation; Labex ACTION program (contract ANR-11-LABX-0001-01); Horiba Medical (Dat@diag); the European Research Council (ERC) under the European Union's Horizon 2020 research and innovation programme (grant agreement No. 740355). K.K. has received funding from the European Union's Horizon 2020 research and innovation programme under the Marie-Sklodowska-Curie grant agreement No. GA-2015-713694 ("BECLEAN" project).